# Edge-Dependent Superconductivity in Twisted Bismuth Bilayers

Isaías Rodríguez[1], Renela M. Valladares[2]*, Alexander Valladares[2], David Hinojosa-Romero[2], Flor B. Quiroga[1], and Ariel A. Valladares[1].

[1]Instituto de Investigaciones en Materiales, Universidad Nacional Autónoma de México, Apartado Postal 70-360, Ciudad Universitaria, CDMX, 04510, México.

[2]Facultad de Ciencias, Universidad Nacional Autónoma de México, Apartado Postal 70-542, Ciudad Universitaria, CDMX, 04510, México.

*Corresponding Author: Renela M. Valladares, renelavalladares@gmail.com

## ABSTRACT

Twisted bilayers offer a compelling and, at times, confounding platform for the engineering of new twistronic materials. Whereas standard studies almost exclusively focus on the explicit enigma that is presented by twist-angles, (perhaps better epitomized by the related phenomena that have been observed in twisted bilayer graphene), functional devices necessarily face a fundamental concern: boundary heterogeneity in their structures. In this study, we address this concern by strictly investigating the electronic properties of twisted bismuth bilayers at the flake's edges and the vibrational properties of the flake. Twisted flakes exhibit continuous variations of these properties, away from the bulk, as we herein report using *ab initio* density functional theory, by systematically mapping the drastic evolution of band topology, electronic density of states and possible superconductivity. Our work reveals a dramatic, non-fortuitous consequence of the structural disorder at the edges of the flakes: an enhanced electronic density of states at the Fermi level. This enhancement reaches a maximum of 10 times that of perfect-crystalline bismuth. Given that the superconducting critical temperature, $T_c$, is exponentially dependent on the electronic density of states at the Fermi level, this substantial structural variation immediately suggests a powerful mechanism for vastly increasing $T_c$. We also identify the twist-angle as a new critical parameter in designing novel engineering devices with topologically enhanced properties. Our results provide a necessary theoretical framework for interpreting new data for the upcoming generation of twistronic heterogeneous materials, and paves the way to search for atomic disordered metastable structures that could lead to enhanced superconducting transition temperatures.

## 1. INTRODUCTION

Monolayer bismuthene, defined as a single-atom-jagged allotrope of bismuth, already stands as a remarkable quantum material because of its bandgap (0.5 eV) [1] and intensely strong spin-orbit coupling. Beyond theoretical fascination, these monolayers foster a potentially room-temperature quantum spin Hall effect [2] and the existence of stable disordered edge states that have been experimentally reported [3]. Nevertheless, the ultimate challenge in modern material science is not a single layer's capabilities, but how we can possibly design a composite structure by stacking these monolayers.

The groundbreaking discovery of twisted graphene bilayer (TGB) unveiled the fact that double-layer stacking introduces an extra degree of freedom, the twist-angle; that radically changes the electronic behavior of the material. Unexpected results, from superconductors to related insulators, are obtained by this interlayer coupling that gives rise to flat electronic bands at specific "magic" twist-angles (~1.1°) [3–6], by moiré patterns [4]. However, the discipline faces a major obstacle: whereas early theoretical models employed ideal, unflawed periodicity [7–11], experimental results reveal a natural spatial heterogeneity in the flake's edges [3,12].

In our recent work, for flawless flakes [13], we depicted the deconstruction of the perfect image of twisted bismuth bilayer (TBB). Mere achievement of global twist-angle homogeneity in a finite flake is insufficient to guarantee electronic homogeneity on the whole flake, due to the presence of border effects.

Atomistic simulations unequivocally demonstrate that local lattice relaxation and unavoidable strain gradients cause moiré period distortions, particularly near the flake boundaries. These topological effects are not artifacts; they are stable within created structures, forcing the inevitable conclusion that topological edge reconstruction plays a crucial role in device performance. We have to address this real-world structural disorder so we can actually engineer the electronic usefulness of these systems.

Despite previous results, still lurking at its heart is a persistent inhomogeneity: How accurately do the electronic states evolve at the TBB flake edges from its homogeneous bulk-like configuration as the flake relaxes and the twist-angle gradually increases? This local inhomogeneity, due to the generated topological disorder, is no mere casual defect; it is the generic structural issue of real devices [13–16]. Experiments have already detected anomalous electronic signatures, including pseudogap suppression and the appearance of 1D channels [18], but there is still a missing link: a theory that connects these electronic phenomena directly with disorder and with the twist-angle.

Here, we bridge this critical gap by systematically investigating the electronic properties of the TBB core and edges under selected twist-angle variations. We use our well-established *ab initio* methodology, based on the Density Functional Theory [19], to map the evolution of band topology, the electronic density of states (eDoS, N(E)), and related properties across these distinctive structures. Our results deliver the key finding: a significant enhancement of the eDoS at the Fermi level ($N(E_F)$) due to the flake borders, showing a massive increase of 10 times that of perfect crystalline bismuth. This large increment in the $N(E_F)$ dictates a potent mechanism for achieving a significantly increased superconducting critical temperature ($T_c$) by searching for disordered metastable atomic structures.

## 2. METHODS

We began with the crystalline Bi-I structure [20], subjecting it to a substantial 20×16×2 expansion to yield a massive, yet necessary, 1,160-atom supercell **(Figure 1a)**. This large scale was crucial to accurately capture the physics of the long-range moiré pattern and local twist-angle variations. From this initial supercell, we removed atoms to isolate two freestanding flakes. Each of these flakes contains 145 atoms, resulting in a non-optimized hexagonal structure containing 290-atom system **(Figure 1b)**. Then the flake was geometry relaxed, and a total of thirty-three distinct moiré configurations were prepared. Each one was generated by rotating one flake relative to the other around the normal axis. The specific twist angles selected were 0.0° **(Figure 1c)**, 0.25°, 0.5°, 0.75°, 1.0°, 1.5°, 2.0°, 2.5°, 3.0°, 3.5°, 4.0°, 5.0°, 6.0°, 7.0°, 8.0° **(Figure 1d)**, 9.0°, 9.5°, 10.0°, 10.5°, 11.0°, 12.0°, 13.0°, 14.0°, 15.0°, 16.0°, 17.0°, 18.0°, 19.0°, 19.5°, 20.0°, 20.5°, 21.5°, 22.0°, and 30.0°.

The selected strategy focused on the 0° to 20.0° interval as the primary region for investigation. Within this region, we sampled with high-resolution angles between 0.0° and 1.0° (using 0.25° increments) to clearly register the subtle early moiré pattern onset and potential low-angle effects. Above 1.0°, sampling density was reduced but selectively increased with supplemental points clustered about regions of interest, namely the 1-4° span (1.0°, 1.5°, 2.0°, 2.5°, 3.0°, 3.5°, 4.0°), the 9-11° span (9.0°, 9.5°, 10.0°, 10.5°, 11.0°), and the 19-21° span (19.0°, 19.5°, 20.0°, 20.5°, 21.0°). This targeted densification aimed to investigate potential local maxima in eDoS. Finally, extended coverage up to 30.0° was included to capture the asymptotic behavior as the twist-angle increases. This combined approach of focused densification around points of interest and extended coverage ensured a thorough and efficient characterization of the twist-angle dependence.

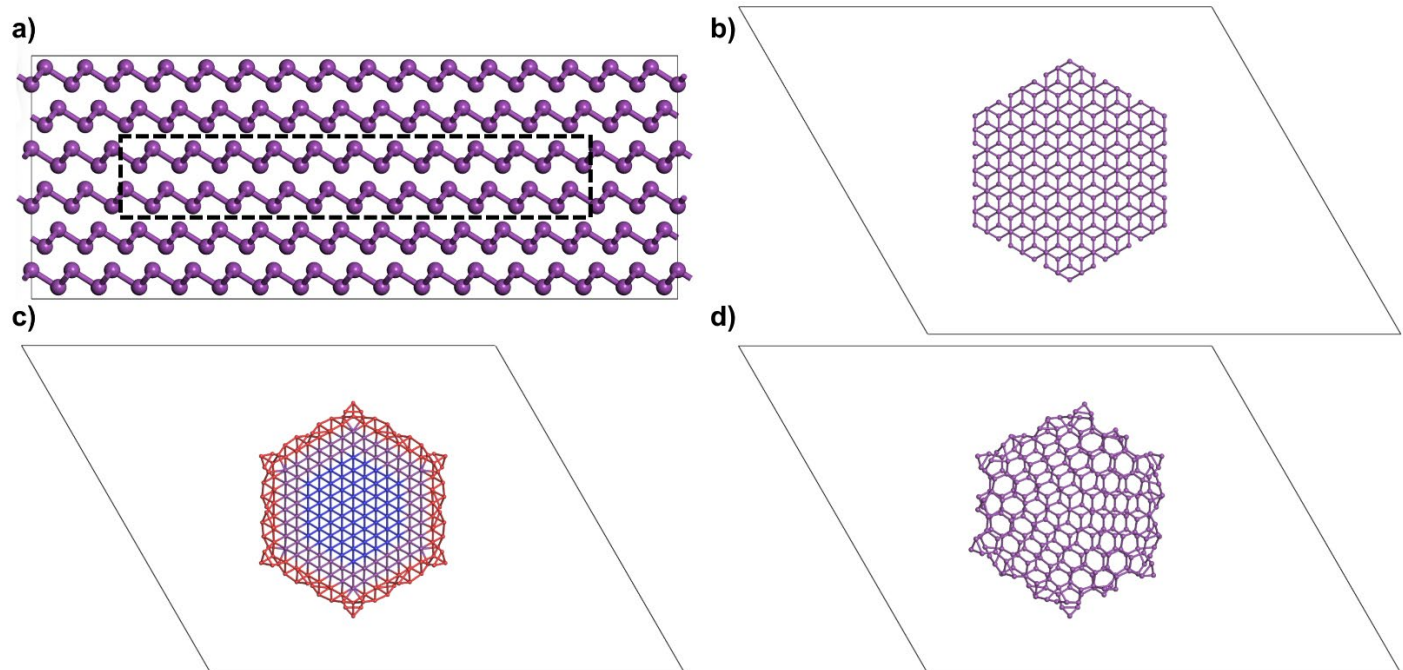

**Figure 1.** Atomistic models of bilayer bismuth flakes. **a)** A ZY plane view of the 20×16×2 supercell of Bi-I used as the starting structure. The broken lines indicate one flake. **b)** An XY plane view of the 290-atom for the non-rotated and non-optimized bilayer bismuth flake. **c)** An XY plane view of the optimized, non-rotated bismuth bilayer flake. The flake's core in blue, the flake's edge in red. **d)** An XY plane view of the optimized structure of the flake with a rotation of 8.0° between layers.

The geometry optimization (GO) process was conducted using the Broyden-Fletcher-Goldfarb-Shanno (BFGS) algorithm [21-24]. The convergence thresholds were $10^{-5}$ Ha for energy, 0.002 Ha/Å for maximum force, and 0.005 Å for maximum displacement. During relaxation, rotation axes were held fixed along the z-axis to conserve the desired twist-angles. Thermal stability was verified through harmonic vibrational analysis, confirming the nonexistence of imaginary frequencies above 10 cm$^{-1}$. The optimized structure of the TBB with a rotation of 8.0° between layers can be seen in **Figure 1d**.

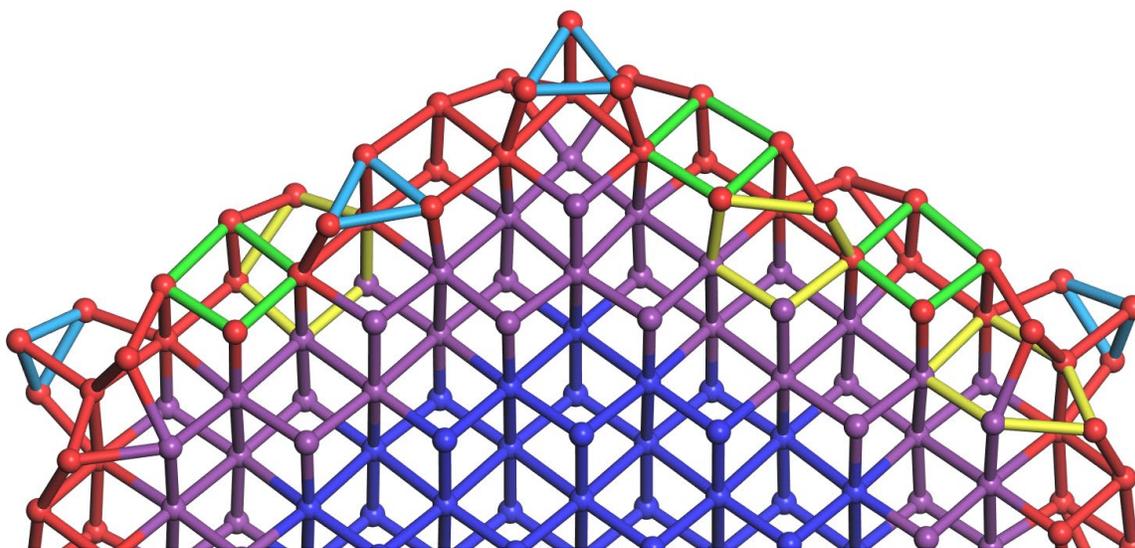

Figure 2. Optimized atomic structure of the flake. Core and edge atoms are represented by blue and red spheres, respectively. Colored bonds indicate specific geometric motifs: triangular (cyan), square (green), and pentagonal (yellow).

Each of the resulting structures is composed of three distinct regions, as illustrated in **Figure 2**. At the center is a crystalline-like core with a diameter of approximately 17 Å (~10 atomic layers), represented by blue atoms. The outermost region, surrounding the core, is a disordered edge that is ~5 Å thick (~3 atomic layers), shown as red atoms. Separating these two is a transition region of at least two atomic layers (purple atoms) that is neither fully crystalline nor disordered. This composite structure results in an overall flake geometry with a maximum and minimum diameter of 40.98 Å and 34.97 Å, respectively. To visualize the local topology within these regions, the bonding network is color-coded: cyan linkages indicate triangular motifs, green bonds highlight quadrilateral motifs (deformed squares) and yellow bonds indicate deformed pentagonal rings.

Electronic structure calculations were performed using DMol³ within the Materials Studio Suite [25]. Parameters adopted for the density functional theory consisted of the Vosko-Wilk-Nusair exchange-correlation functional (VWN-LDA) [26], a double numerical plus polarization basis set, along with a density functional semi-core pseudopotential (DSPP) treatment for the cores. The calculations used a Γ-centered 4×4×1 k-point sampling, a 6.0 Å real-space orbital cutoff for the wave-function basis set, and a $10^{-6}$ Ha energy tolerance.

In addition, we used an integration grid with octupolar angular momentum fitting functions. Lastly, the eDoS was obtained by means of single-point energy calculations on optimized structures. Vibrational densities of states (vDoS, $F(\omega)$) were calculated using the finite displacement method with a 0.002 Å displacement step to construct the dynamical matrix.

3. RESULTS AND DISCUSSION
    3.1. Structural analysis

Characterization of the bismuthene structures was accomplished through extensive study of the optimized flakes' geometry. ***Correlation*** [27], a computer code developed by our group, was used for this purpose. This tool enabled the precise calculation of two key functions for structural description: the pair distribution function (PDF, g(r)), and the plane-angle distribution function (F(θ)).

The pair distribution function, gives information about the probability of finding an atom at a distance r away from a selected reference atom, providing a quantitative description of the atomic density as a function of distance. The plane-angle distribution function, on the other hand, is crucial for understanding the orientation of bonds and the local arrangement within the structure, as it quantifies the distribution of angles formed by neighboring atoms. The results obtained using this analysis are discussed in what follows.

Following the GO process, the flake exhibits two distinct regional behaviors. The core of the flake undergoes a rigid-like compression relative to the bulk Bi-I crystal structure (**Figure 1a**), an effect that is manifested in the comparison of the PDF in **Figure 3a**. A 3.4% compression in volume of the parent Bi-I structure produces an exact replica of first and second neighbor interatomic distances in the flake´s core atoms. The third and fourth nearest-neighbor interatomic distances are also compressed and matched with an error less than 4%, as depicted in **Figure 3b**.

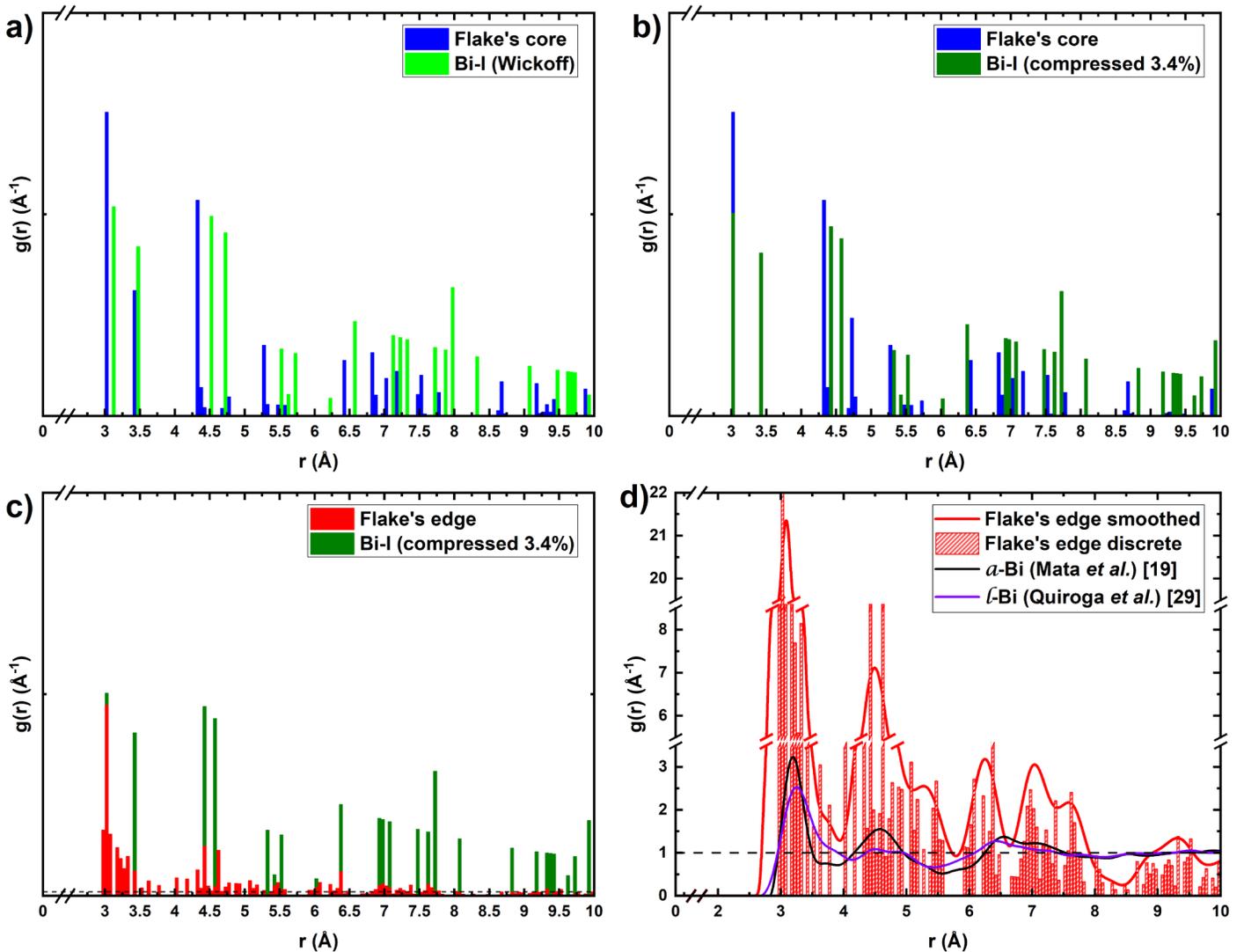

**Figure 3.** Pair Distribution Functions of the bilayer bismuth flakes. **a)** PDF of the flake's core atoms compared to the bulk Bismuth I phase, highlighting the crystalline nature of the core. **b)** Comparison of the flake's core PDF with that of a 3.4% compressed bulk Bismuth I structure, revealing the core's compressed state. **c)** PDF of the flake's edge atoms compared to compressed Bismuth I, showing significant deviation from the bulk crystal structure. **d)** Smoothed and discrete PDF of the flake's edge (in red), compared with amorphous (in black) [19,28] and liquid bismuth (in purple) [29], to illustrate the disordered nature of the flake's edge.

Optimization led to a result where the flake's edge became heavily disordered, almost amorphous-like (**Figure 1d**), a state clearly characterized by its PDF (**Figure 3d**), the edge PDF deviates significantly from the crystalline structure. Then we compared the flake's edge PDF with that of amorphous and liquid bismuth [19,28, 29] as shown in **Figure 3d,** finding wide delocalized peaks characteristic of disordered materials. This finding provides compelling evidence for the disordered nature of the flake's edge. A further analysis reveals that the flake's edge PDF lacks the characteristic shoulder between 3.5 Å and 4.0 Å, which is a hallmark of liquid bismuth [26], definitively ruling out a molten state. Furthermore, the edge of the flake shows a 3.9% compression compared to both the amorphous and liquid bismuth, suggesting a unique and highly strained structural arrangement.

Pair distribution functions have been shown to be effective in the description of significant structural variations across the bismuth flake; the core's compression and the edge's disordering, provides information only about interatomic distances. To gain a more complete picture of the local atomic arrangement and validate these findings, it is essential to analyze the bond angles of the geometry relaxed structures. In **Figure 4**, we show the results of $F(\theta)$ for the untwisted flake (0.0° twist-angle).

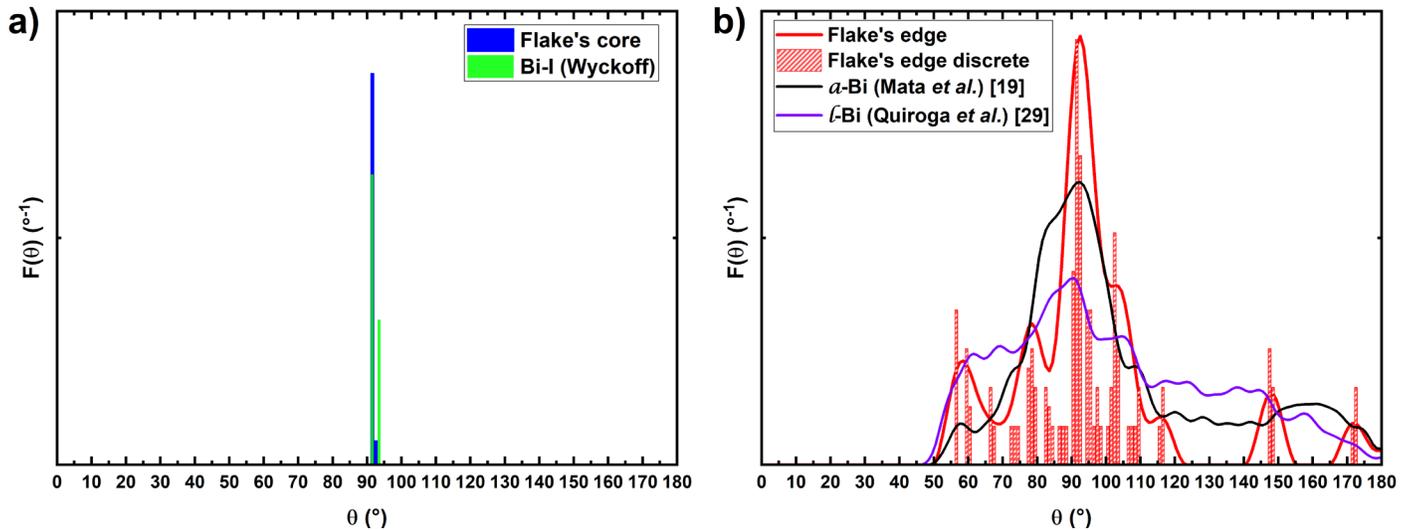

**Figure 4.** Plane-Angle Distribution (F(θ)) analysis for 0.0° twist-angle. **a)** F(θ) of the flake's core (blue) contrasted with characteristic peaks of crystalline Bismuth I (green), demonstrating the preservation of the core's ordered structure. **b)** Smoothed and discrete F(θ) of the flake's edge (red) compared to amorphous (black) [19,28] and liquid bismuth (purple) [29]. The broad, featureless distribution of the edge confirms its disordered, non-crystalline state.

Analysis of the F(θ), reveals distinct differences between the flake's core and edge. As shown in **Figure 4a**, the flake's core F(θ) exhibits two peaks at 91.5° and 92.5°, closely mirroring the crystalline Bismuth I structure's main peak at 91.5° and secondary peak at 93.5°. This close correspondence confirms that the core preserves the ordered, symmetrical lattice of the parent Bi-I crystal. In stark contrast, the flake's edge presents a broad, prominent peak at 90° and a secondary broad peak at 57° (**Figure 3b**). This multimodal distribution indicates a breakup of the hexagonal lattice into a random network of deformed triangles (57°), squares (90°), and pentagons (116°). Such peak broadening and geometric diversity confirm the disordered nature of the edge, aligning with the PDFs results.

### 3.2. Electronic density of states

The electronic density of states for the TBB samples reveals a complex structure comprised of two major energy bands (**Figure 5**). The first band, spanning from approximately −15 eV to −8 eV, is primarily composed of deeper-lying Bi 6s and 6p states. Features within this region, such as the distinct peaks around −12 eV and −10 eV, are attributed to localized atomic orbitals. In **Figure 5**, we can observe the variation across all studied twist angles, both in the flake's core and in the flake's edge.

The second band, narrower, expands from approximately −5 eV to the Fermi level ($E_F$), and contains the electronic states crucial for this material's electronic properties. This region is highly sensitive to the interlayer twist-angle in the flake's edge region (**Figure 5e-h**) and is dominated by Bi 6p orbitals in the whole flake (**Figure 6**). The non-zero electronic density of states at the Fermi level, $N(E_F)$, indicates a metallic-like character. These two primary bands are separated by a wide energy gap between -8 eV and -5 eV.

The significant difference in structural topology and confinement between the flake's core and the flake's edge requires analyzing these regions separately to understand the electronic properties of the system.

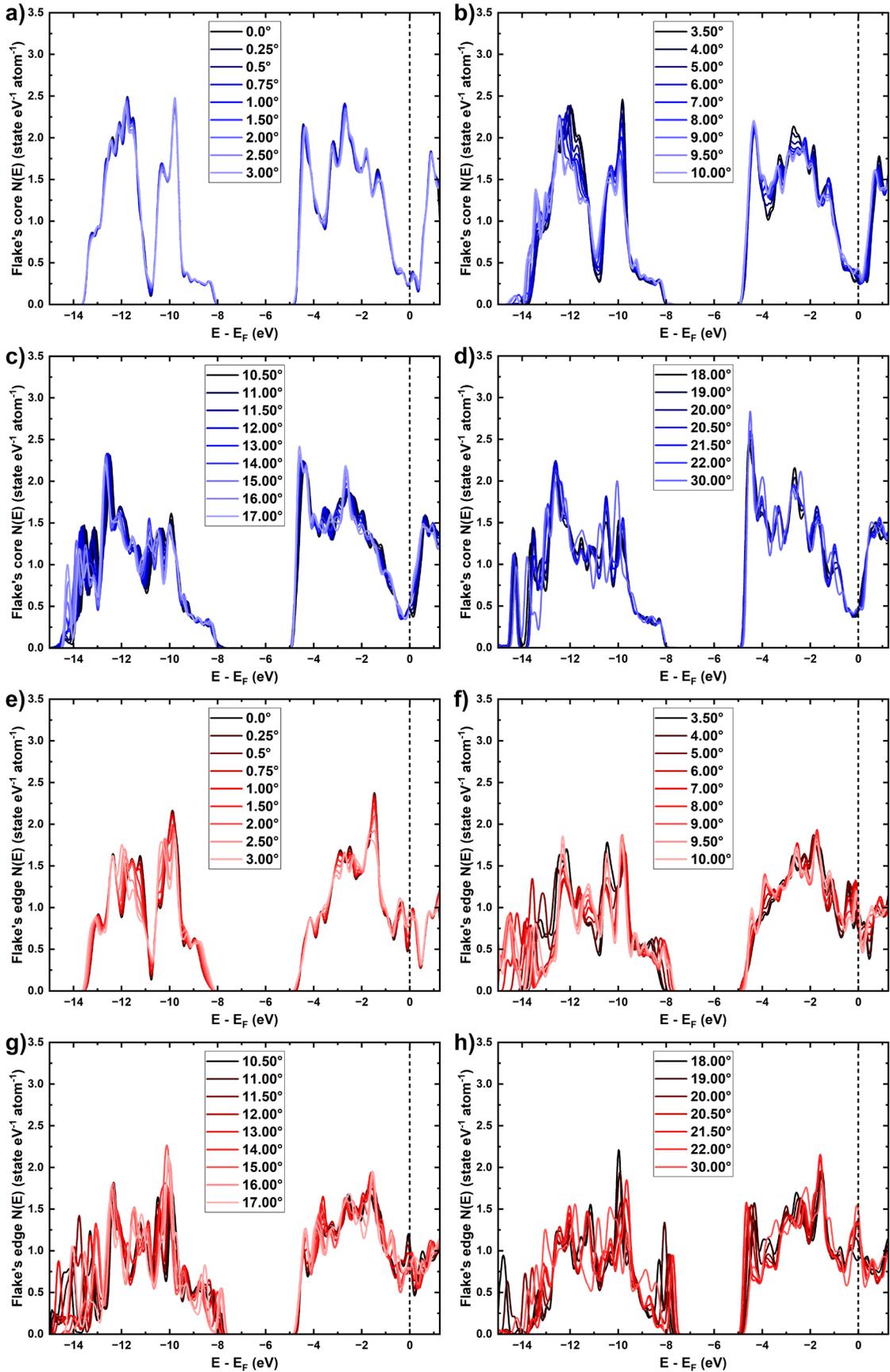

**Figure 5**. Electronic density of states for the Twisted Bismuth Bilayers. Panels **a)** to **d)** show the eDoS calculated for the core region of the flake, grouped by angular range: **a)** $0.0° \leq \theta \leq 3.0°$, **b)** $3.5° \leq \theta \leq 10.0°$, **c)** $10.5° \leq \theta \leq 17.0°$, and **d)** $18.0° \leq \theta \leq 30.0°$. Panels **e)** to **h)** present the corresponding eDoS for the edge region of the flake, using the same angular grouping as **a)**-**d)**.

Analysis of the flake's core region (**Figure 5a-d**), reveals a surprising twist-angle invariance. The electronic structure is dominated by a non-zero valley centered at the Fermi level, confirming the semi-metallic character of the core region. Crucially, the magnitude of N($E_F$), shows negligible variation across the entire range of the studied angles (**Figure 7**). This invariance, is observed even around the previously reported small magic-angle of 0.5° [13], leading us to conclude that the states contributing to the Fermi surface of the flake's core are immune to slight variations in the moiré patterns for the entire range.

The electronic density of states of the flake's edge (**Figure 5e-h**) presents a different picture; it is characterized by the presence of a sharp peak at the Fermi level, resembling a van Hove singularity (vHs). The presence of this peak of the flake's edge, a feature typically associated with energy localized electronic states, indicates that the structural disorder, inherent to the flake boundary reconfigures the electronic states in the vicinity of the Fermi level. This boundary effect perturbs and localizes the electronic states at the Fermi level.

The persistence of a non-zero N($E_F$) in both the core and edge regions confirms that the system metallic-like character is maintained throughout the whole flake, although the energy states distribution near $E_F$ is profoundly modulated by the local geometric edge structure.

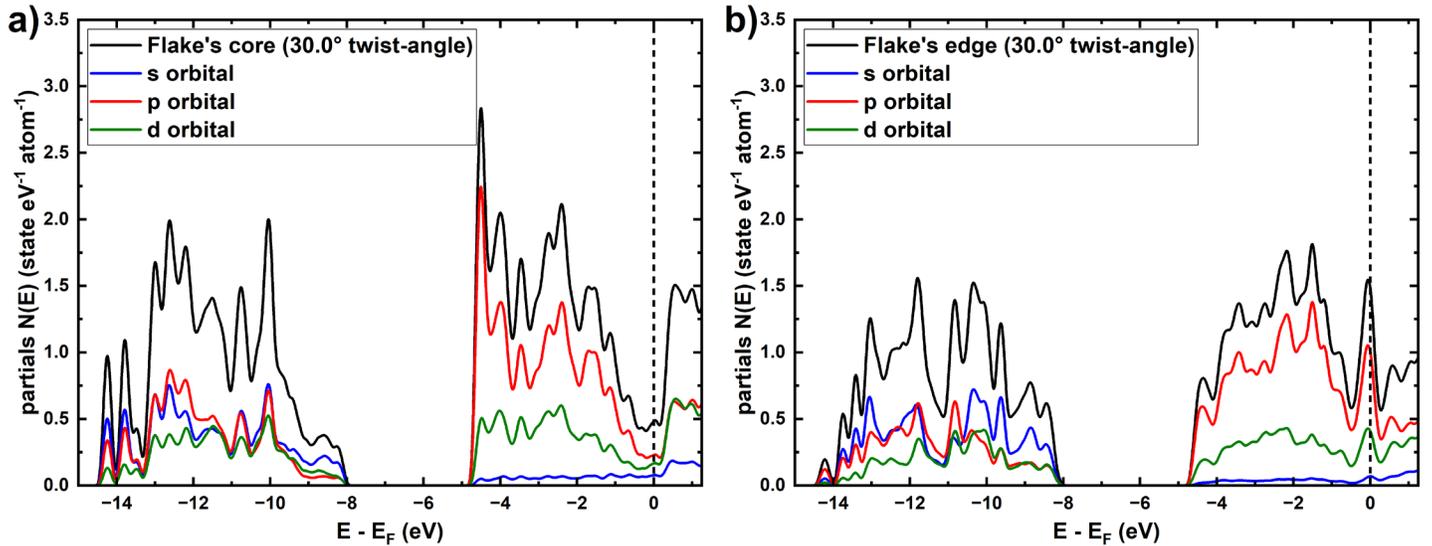

**Figure 6**. Partial and total electronic density of states for the twisted bismuth bilayer at a twist-angle of 30°. **a)** N(E) for the flake's core. **b)** N(E) for the flake's edge. The total N(E) is shown in black. Orbital-projected contributions from the Bi 6s, 6p, and 6d orbitals are displayed in blue, red, and green, respectively.

This pronounced structural and electronic dichotomy between the flake regions is quantitatively reflected in the N($E_F$) as a function of the interlayer twist-angle, presented in **Figure 7**. The N($E_F$) of the flake's core (blue squares) displays a relatively invariant bulk-like structure, peaking broadly at 0.59 states·eV$^{-1}$·atom$^{-1}$ for θ=15.0°. This electronic behavior for N(E) of the core is largely insensitive to the twist-angle, which strongly suggests that the flake's edge region is primarily responsible for the system sensitive electronic phenomena, including the previously reported superconducting behavior effect at 0.5° for pristine TBB [13].

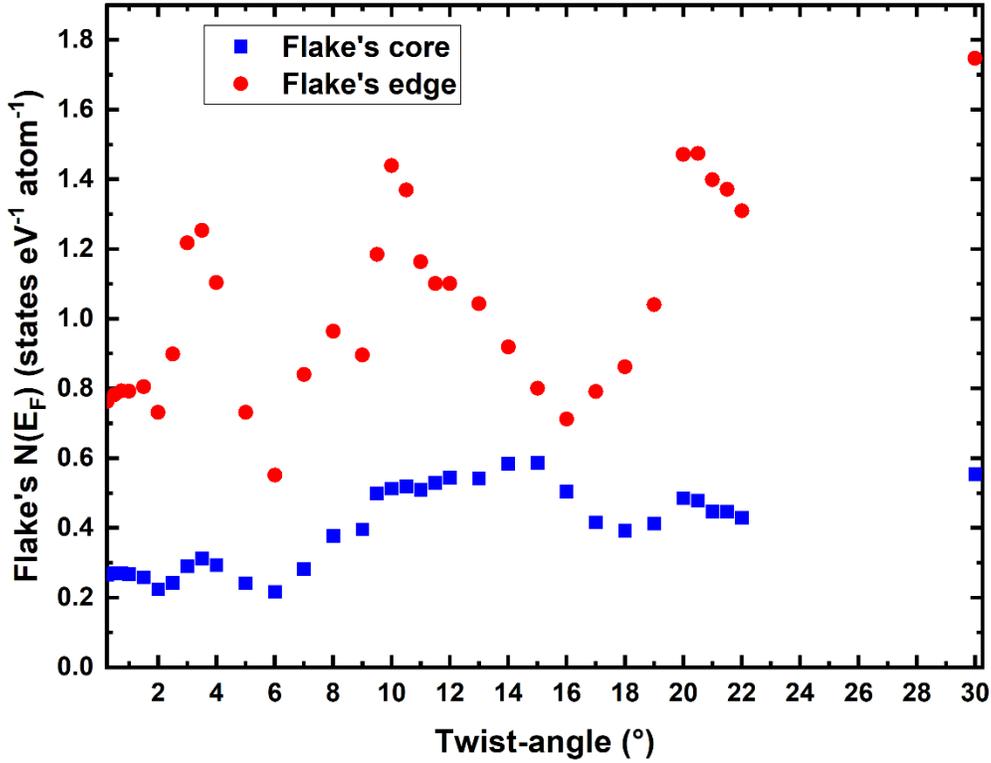

**Figure 7.** The electronic density of states at the Fermi level, reveals a striking contrast between the flake's core (blue squares) and its edge (red circles). The core exhibits a relatively invariant, bulk-like behavior, with a single broad maximum of 0.59 states·eV$^{-1}$·atom$^{-1}$ at θ=15.0°. In contrast, the edge displays a highly oscillatory, non-monotonic N($E_F$) across the studied range, featuring four distinct maxima at 3.5°, 10.0°, 20.5°, and 30.0°.

In sharp contrast, the N($E_F$) for the flake's edge (red circles) is highly sensitive to the rotation, showing a non-monotonic, oscillatory, dependence on the twist-angle, featuring four distinct local maxima at 3.5°, 10.0°, 20.5°, and 30.0°, representing a massive 10-fold increase over the N($E_F$) of perfect crystalline Bi-I structure, 0.15 states·eV$^{-1}$·atom$^{-1}$ [13, 19] This behavior demonstrates that the edge's electronic properties are decoupled from the TBB core, and are instead, dominated by the structural relaxations and the boundary states.

### 3.3. Superconductivity and Debye temperatures

Investigations of twisted van der Waals heterostructures, a field of research termed as twistronics, have effectively transformed the condensed matter physics dynamics ever since the magic-angle TGB phenomenon of superconductivity was discovered in 2018 [30]. The foundational principle is that a rotational misalignment between atomic layers generates moiré patterns which dramatically modifies the electronic band structure. In particular, this situation can give rise to the existence of quasi-flat bands, where the electron group velocity is near zero [31]. Flat bands energy-localize the electronic states, resulting in an extremely high electronic density of states around the Fermi level and enhancing electron-phonon interaction effects.

For the current work on the TBB, the high value of N($E_F$) established in the previous section, offers the onset condition for a superconducting phase. The following results detail the calculation of the Debye temperature, $\theta_D$, as well as the $T_c$, and explore the profound influence of the interlayer twist-angle as well as the structure defects on the flake's edge *versus* the core.

The initial step in the estimation of the hypothetical T_c is to introduce the lattice dynamics to analyze the electron-phonon interaction. For the present work the $F(\omega)$ was calculated using the Finite Displacement Method (FDM) (often referred to as the frozen-phonon method). The same electronic parameters used for the N(E) calculation were employed for the $F(\omega)$. To ensure computational stability and accuracy in obtaining the Hessian matrix, a finite difference displacement of 0.005 Å was used. Due to the supercell nature of the calculation in the FDM, the $F(\omega)$ represents the q=0 (Gamma point) phonons. Subsequently, the Debye frequency ($\omega_D$) and the Debye temperature ($\theta_D$) were calculated from the resulting $F(\omega)$ using the formula proposed by Grimvall [32]:

$$\omega_D = exp\left(\frac{1}{3} + \frac{\int_0^{\omega_{max}} ln(\omega)F(\omega)d\omega}{\int_0^{\omega_{max}} F(\omega)d\omega}\right),$$

and

$$\theta_D = \frac{\hbar\omega_D}{K_B},$$

where $\hbar$ is the reduced Planck constant, and $K_B$ is the Boltzmann constant.

The calculated $\theta_D$, a crucial parameter governing the electron-phonon interaction, exhibits a non-monotonic dependence on the interlayer twist-angle, as illustrated in **Figure 8**. Within the measured range of 0.0° to 10.0°, $\theta_D$ shows a periodic variation. The largest calculated value was 102.99 K for a twist-angle of 2.0°, and the lowest was 96.520 K for a twist-angle of 9.0°. This angular dependence is significant, suggesting that the moiré pattern strongly modulates the lattice's vibrational properties.

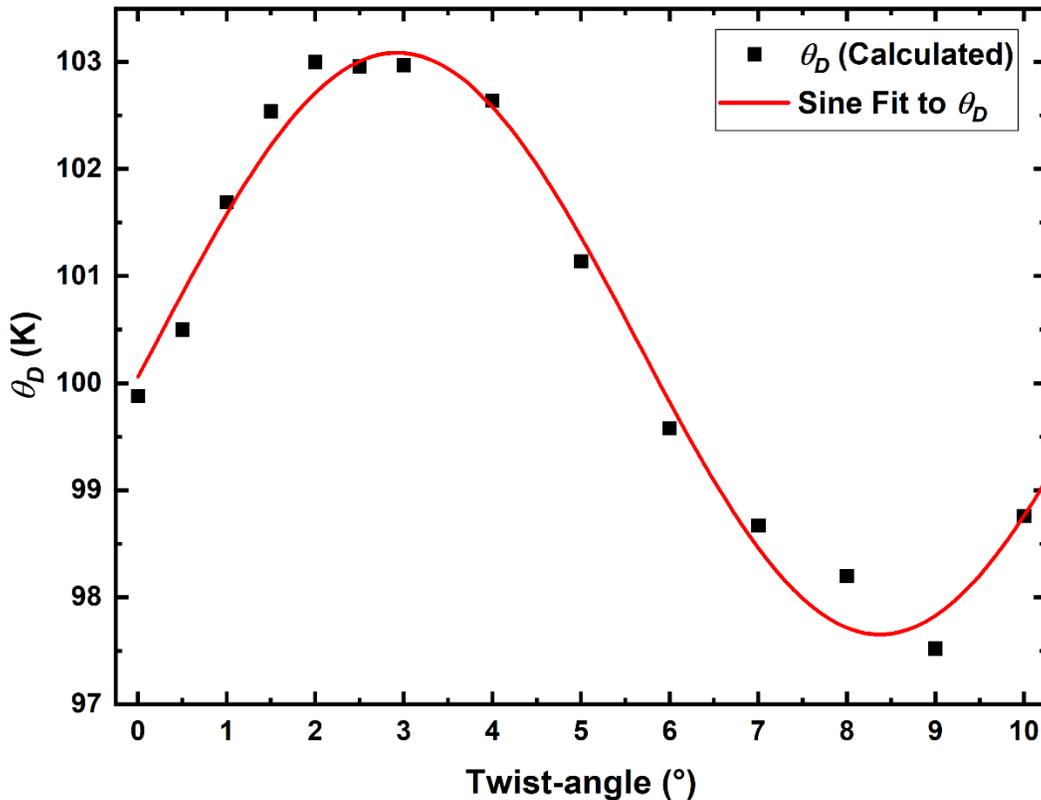

Figure 8. Debye temperature ($\theta_D$) (in Kelvin) for the twisted bismuth bilayer system, calculated from the vibrational density of states via the Grimvall method, shown for twist angles ranging from 0.0° to 10.0°. The calculated values are represented by black squares. The data shows a periodic oscillation, which is well-described by the sine fit (red line), highlighting the strong, non-monotonic influence of the moiré patterns on the Debye temperature.

To predict $\theta_D$ across the full angular spectrum and to quantify this periodic relationship, the data was successfully modeled using a sine fitting function. The resulting fit provided excellent statistical agreement, yielding an $R^2$ value of 0.9831 and a $\chi^2$ value of 0.08445. The functional form of the fit is:

$$\theta_D = 100.37 + 2.72\, sin\left(\frac{\pi(\theta - 0.2°)}{5.45°}\right) \text{ K}.$$

This oscillation in $\theta_D$, when coupled with the calculated N($E_F$) variation (Section 3.2), leads to an angle-dependent relation of both the electron-phonon coupling and the resulting superconducting transition temperature.

In order to estimate the superconducting transition temperature, we assume that the Debye temperature has the same value at any given twist-angle for the crystalline-like flake's core and the disordered-like flake's edge. This approach is justified by comparing several allotropes of bismuth. The $\theta_D$ of bulk crystalline bismuth Bi-I is 112 K, Bi-II is 115.5 K, Bi-III is 96.4 K and Bi-IV is 102.1 K [33–36], while for amorphous bismuth it has a value of 100 K [19,37]. This small difference, between several different structures, suggests that $\theta_D$ is not highly sensitive to the changes in atomic structure. Since the primary distinction between our flake's core and flake's edge is the loss of translational symmetry rather than local bond strength (compression of 4% for the core and 3.9% for the edge), we posit that their intrinsic $\theta_D$ values should be similarly close. Therefore, for subsequent calculations of the $T_c$, the same angle-dependent $\theta_D$ derived above, is applied to both the crystalline-like core and the disordered-like edge regions.

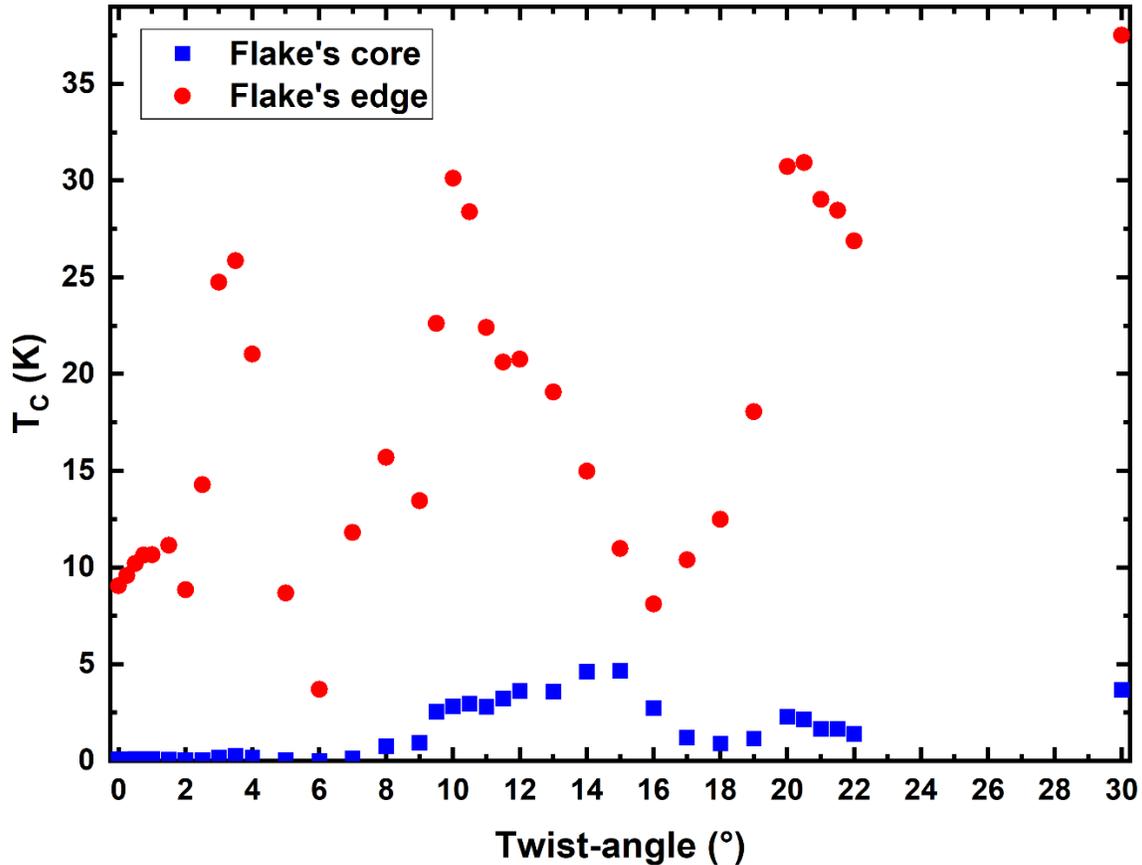

**Figure 9**. Calculated superconducting critical temperature, $T_c$, for the twisted bismuth bilayer system, comparing the flake's core (blue) and the flake's edge (red) regions across the full angular range. The $T_c$ values for the core follows a smooth trend for all range of studied twist-angles. In contrast, the edge region exhibits a highly oscillatory $T_c$, culminating in a remarkable maximum value of 37.53 K at a twist angle of 30.0°.

In order to quantitatively estimate the $T_c$ of the TBB system, we employ the Mata-Valladares approach [19]. It is a well-proven method for the $T_c$ estimation from the calculated electronic and vibrational properties. We selected this approach based on its past success estimating and predicting $T_c$ for various allotropes and phases of bismuth [19,35,38–40]. This approach takes the $N(E_F)$ values obtained in Section 3.2 and the angle-dependent $\theta_D$ values obtained in this section, and 0.53 mK, the experimental value of $T_c$ for Bi-I [41], to compute the superconducting transition. The resulting $T_c$s are shown in **Figure 9**. The calculated $T_c$ shows a very large contrast between the edge and core regions of the flake, as anticipated in view of the electronic structure behavior obtained for the $N(E)$ (**Figure 7**).

The $T_c$ for the flake's core manifests a relatively bounded dependence on the twist-angle, reaching a maximum value of 4.67 K at a twist-angle of 15.0°. This observed behavior fits well with the established model of phonon-mediated (BCS-type) superconductivity in crystalline allotropes of bismuth [42–44].

In contrast, the $T_c$ of the edge of the flake shows a strong oscillatory and non-monotonic character. It exhibits four clear local maxima at precisely the same locations as the peaks in $N(E_F)$ obtained in the edge region analysis (**Figure 7**). Remarkably, the $T_c$ culminates in an interesting and provoking maximum of 37.53 K at twist-angle of 30.0°. As expected, the correlation between the $T_c$ maxima (**Figure 9**) and the $N(E_F)$ peaks (**Figure 7**) strongly suggest a major dependence on the $N(E_F)$, with a minor influence of the variation in $\theta_D$ (**Figure 8**). This strong dependence on the twist-angle rotation reveals that the high $T_c$ in the flake's edge is a result of the strong energy localization of the electronic states in the borders of the flake (i.e., the vHs [45,46]).

Our work identifies two principal mechanisms for enhancing the superconducting critical temperature in twisted bilayer systems: twist-angle engineering and disorder, the latter being a well-documented route to higher $T_c$ [47-49]. This finding is consistent with a growing consensus that for 2D materials disorder can enhance superconductivity [50-54].

4. CONCLUSIONS

This density functional theory work reveals the dominant role of structural heterogeneity in modulating the electronic properties of the twisted bismuth bilayers. Our findings confirm a clear relation between the flake's disorder, the electronic structure, and a potential high temperature superconductor.

The full geometry optimization confirms a profound structural dichotomy within the flake: core *vs* edge. The core maintains a compressed, crystalline-like structure, while the flake's edge undergoes significant relaxation, adopting a disordered-like configuration. This structural contrast is central to the electronic properties of the material.

Because of the structure dichotomy, the electronic density of states at the Fermi level shows a dual behavior. The crystalline-like core maintains a semi-metallic character with $N(E_F)$ showing a minimal variation for the whole range of twist-angles. Its profile is bounded with a peak value of 0.59 states·eV$^{-1}$·atom$^{-1}$ at $\theta$=15.0°. The disordered-like edge, in contrast, is characterized by a highly oscillatory $N(E_F)$ profile, driven by sharp localized peaks in the electronic density of states.

The calculated superconducting critical temperature reveals a stark contrast in behavior. The core exhibits a conventionally low $T_c$, peaking at 4.67 K for $\theta$=15.0°. The edge $T_c$ exhibits a dramatic, non-monotonic trend, reaching an unprecedented maximum of 37.53 K for $\theta$=30.0°. The perfect one-to-one correlation between the $T_c$ and the $N(E_F)$ peaks confirms that the structural disorder at the edge, when used within the BCS framework, is the definitive driver of the extraordinary $T_c$ enhancement.

In summary, this work establishes edge-angle engineering as a critical new design principle for creating superconducting topologically enhanced twisted bilayered devices. The potential for $T_c$ enhancement at the flake's edge by over 7000 times that of 0.53 mK for the crystalline Bi-I, demonstrates that structural defects and edge effects, previously considered parasitic, can be leveraged to drive highly desirable emergent properties. Our results seem to indicate that the generation of new structures either by high pressure, or by twisting bilayer, or by generating new disorder structures, at the borders of flakes, is the way to go to search for new superconductors.

**ACKNOWLEDGEMENTS**
I.R. and D.H.-R. thank SECIHTI and DGAPA-UNAM (PAPIIT) for their respective postdoctoral fellowships. F.B.Q. thanks SECIHTI for her doctoral fellowship. A.A.V., R.M.V., and A.V. acknowledge DGAPA-UNAM (PAPIIT) for continued financial support under Grant No. IN118223. This research was also supported by SECIHTI under Grant No. CBF-2025-G-886. Computational resources were partially provided by the Supercomputing Center of DGTIC-UNAM through the project LANCAD-UNAM-DGTIC-131. The authors are grateful to María Teresa Vázquez and Oralia Jiménez for their assistance with information retrieval and to Alejandro Pompa for his technical support and maintenance of the computing site at IIM-UNAM.

**AUTHOR CONTRIBUTIONS**
This research was conceived and designed by I.R. and A.A.V., with input from R.M.V., A.V., D.H.-R. and F.B.Q. All simulations were performed by I.R. All authors contributed to the discussion and analysis of the results. I.R. and A.A.V. wrote the initial draft of the manuscript, which was subsequently enriched and approved by all co-authors for publication.

**COMPETING INTERESTS STATEMENT**
The authors declare no conflict of interest in this work.

**DATA AVAILABILITY STATEMENT**
The datasets used and analyzed during the current study are available from the corresponding author on reasonable request.